\documentclass[epj]{svjour}
\usepackage{graphics}
\usepackage{color}
\usepackage{graphicx}
\usepackage{dcolumn} 
\usepackage{bm}  

\begin{document}
\title{Symmetry energy from the nuclear collective motion: constraints from dipole,
quadrupole, monopole and spin-dipole resonances}
\author{G. Col\`o\inst{1,2}, U. Garg\inst{3,4}, H. Sagawa\inst{5,6}
}                     
\institute{Dipartimento di Fisica, Universit\`a degli Studi di Milano, via Celoria 16, I-20133 Milano, Italy \and INFN, sezione di Milano, via Celoria 16, I-20133 Milano, Italy \and
Physics Department, University of Notre Dame, Notre Dame, Indiana 46556, USA \and Joint Institute for Nuclear Astrophysics, University of Notre Dame, Notre Dame, Indiana 46556, USA \and
Center for Mathematics and Physics, University of Aizu, Aizu-Wakamatsu, Fukushima 965-8560, Japan \and RIKEN Nishina Center, Wako 351-0198, Japan}
%
%
%
\date{Received: date / Revised version: date}
%
\abstract{
The experimental and theoretical studies of Giant Resonances, or more generally of the nuclear collective vibrations, are a 
well established domain in which sophisticated techniques have been introduced and firm conclusions reached after an effort of several
decades. From it, information on the nuclear equation of state can be extracted, albeit not far from usual nuclear densities.
In this contribution, which complements other contributions appearing in the current volume, we survey some of the constraints
that have been extracted recently concerning the parameters of the nuclear symmetry energy. Isovector modes, in which neutrons
and protons are in opposite phase, are a natural source of information and we illustrate the values of symmetry energy around
saturation deduced from isovector dipole and isovector quadrupole states. The isotopic dependence of the isoscalar monopole energy
has also been suggested to provide a connection to the symmetry energy: relevant theoretical arguments and experimental results are thoroughly
discussed. Finally, we consider the case of the charge-exchange spin-dipole excitations in which the sum rule associated with
the total strength gives in principle access to the neutron skin and thus, indirectly, to the symmetry energy.  
} 

\authorrunning{G. Col\`o, U. Garg, H. Sagawa}
\titlerunning{Symmetry energy from the nuclear collective motion}

\maketitle

\section{Introduction}\label{intro}

As is testified by the variety of contributions in this volume, complementarity
of the sources of information is a vital component of our understanding of the symmetry
energy. In our contribution, we review several attempts to use the nuclear
collective excitations as a tool to infer the properties of the symmetry energy.
We stress that in most of these cases, although we can only access 
densities that are relatively close to the usual nuclear density, the information can be
considered as quite accurate due both to well established experimental techniques and to the availability of
microscopic methods that have been tested against many other observables.
This is at variance with other situations (astrophysical observations and, to some
extent, heavy-ion collisions) in which one is potentially able to explore a broader
range of densities, but at the expense of facing with more global and specific
uncertainties.

We start by reviewing the basic equations related to symmetry energy. 
For any nuclear system the total energy must depend both on neutron and proton 
densities $\rho_n$ and $\rho_p$, 
\begin{equation}\label{edf}
E = \int d^3r\ {\cal E}(\rho_n(\vec r),\rho_p(\vec r)),
\end{equation}
where $\cal E$ is the energy density and we have assumed locality
for the sake of simplicity. In the following, we will use $q$ as a generic label
for neutrons and protons. In finite systems the energy can actually
depend not only on the spatial densities, but also on their 
gradients $\nabla\rho_q$, on the kinetic energy densities
$\tau_q$, as well as on other generalised densities like the spin-orbit 
densities $J_q$; however, in infinite matter, one has a simple expression in terms
of the spatial densities only (cf., e.g., Ref.~\cite{Bender:2003}).

Instead of $\rho_n$ and $\rho_p$, one can use the total density $\rho$ and 
the {\em local} neutron-proton asymmetry,
\begin{equation}
\beta \equiv {\rho_n-\rho_p \over \rho}.
\end{equation}
In asymmetric matter, we can make a further simplification on ${\cal E}(\rho,\beta)$ by
making a Taylor expansion in $\beta$ and retaining only
the quadratic term (odd powers of $\beta$ are obviously forbidden
due to isospin symmetry), 
\begin{eqnarray}\label{def_sym}
{\cal E}(\rho,\beta) & \approx & {\cal E}_0(\rho,\beta=0) +
{\cal E}_{\rm sym}(\rho) \beta^2 \nonumber \\
& = & {\cal E}_0(\rho,\beta=0) +
\rho S(\rho) \beta^2.
\end{eqnarray}

The first term on the r.h.s. is the energy density of
symmetric nuclear matter ${\cal E}_{\rm nm}$, while the
second term defines the main object of all studies in this volume,
namely the {\em symmetry energy} $S(\rho)$. The symmetry
energy at saturation $S(\rho_0)$ is denoted by different
symbols in the literature {\em viz.} $J$, $a_\tau$ or $a_4$; we shall use $J$ in what 
follows. We stress, however, 
that Eq.~(\ref{def_sym}) is not really a simplification: the
coefficient of the term in $\beta^4$ which should follow is negligible in 
most models at the densities of interest for this work 
(see, e.g., Ref. \cite{vidana2009} and in particular Fig. 1 of
that work; this conclusion has been also systematically checked
in Ref. \cite{Trippa:2008} with all the models used therein).
We remind the reader that the pressure can be written
in a uniform system as
\begin{equation}
P = - \left. {\partial E \over \partial V} \right|_A =
\left. \rho^2 {\partial \over \partial \rho}{{\cal E} \over
\rho} \right|_A,  
\end{equation}
where $A$ is the nucleon number.  Thus, although it is customary to refer to the energy per particle as
the ``equation of state'', the relationship with the quantity that
better fits such name, that is, the pressure as a function of
the density, is evident from the latter equation.

The overall trend of the symmetry energy is
poorly known, but the main quantities on which attention
has been focused are 
\begin{eqnarray}\label{parameters}
J & \equiv & S(\rho_0), \nonumber \\
L & \equiv & 3\rho_0\ S^\prime(\rho_0), \nonumber \\
K_{\rm sym} & \equiv & 9\rho_0^2\ S^{\prime\prime}(\rho_0).
\end{eqnarray}
$L$ is often referred to as the ``slope parameter''. 

It is expected, therefore, that the isovector giant resonances may be 
the main source of information for the symmetry energy.
They are collective excitations in which most of the nucleons
participate, as it is known from the fact that they exhaust a
large fraction of the appropriate sum rules. Especially
in heavy nuclei we can assume that if an isovector
external field displaces the protons with respect to the
neutrons, by creating a local proton-neutron 
asymmetry, the restoring force in the harmonic approximation
can be related to
\begin{equation}
\frac{\delta^2 E}{\delta\beta^2}, 
\end{equation}
if the oscillations involve only variations of $\beta$. Of course
such simple argument should be taken with care: the nucleus does
not have uniform values of $\rho$ and $\beta$; isospin is not an
exact quantum number so that isoscalar and isovector oscillations
are not well separated; and, finally, quantum effects like shell structure or
pairing may also spoil the simple classical arguments.
However, we will show in what follows that these warnings, although
manifesting themselves in some error bar that we must attribute 
to our extractions of the symmetry energy, do not prevent at all
deducing values for the symmetry energy around saturation.
We shall also show that isoscalar modes
like the monopole resonance can be somewhat related to the
symmetry energy if one observes this mode along an isotopic
chain.

The outline of the paper is the following. In Secs. \ref{ivgdr}, \ref{pdr} 
and \ref{ivgqr} we discuss
some constraints coming from different isovector states, that is, the well-known
giant dipole resonance, the so-called pygmy resonances, and the
isovector giant quadrupole resonance, respectively. In Sec. \ref{incompr} we provide theoretical
arguments why the isoscalar monopole, when measured in nuclei with
neutron excess, could also provide access to some key parameter 
associated with the symmetry energy; the related experimental data
and quantitative conclusions are drawn in Sec. \ref{isgmr}. 
In Sec. \ref{sdr} we move to the spin-dipole mode excited by 
charge-exchange reactions, namely to the
spin-dipole mode of excitation, whose total strength is related with the neutron
skin; we review the experimental findings and the associated theoretical
analysis.
Finally, we provide
a short summary in Sec. 7. 

\section{Symmetry energy from the IVGDR}\label{ivgdr}

The relationship between the symmetry energy and the most collective 
and well known isovector giant resonance, namely the isovector giant dipole
resonance (IVGDR), can be well elucidated by some macroscopic
model. As it has been done in Ref.~\cite{Trippa:2008}, one can start as a guideline
from the hydrodynamical model of giant resonances, as proposed by E. Lipparini 
and S. Stringari~\cite{Lipparini:1989}. They assume an energy functional (\ref{edf}) 
which is simplified yet sufficiently realistic, solve the macroscopic equations 
for the densities and currents, and extract expressions for the moments $m_1$ 
and $m_{-1}$ associated with an external operator $F$ ($m_k\equiv\int dE\ S(E) 
E^k$ where $S$ is the strength function associated with $F$). The expression 
for $m_1$ is proportional to $(1+\kappa)$, where $\kappa$ is the well-known 
``enhancement factor'' which in the case of Skyrme forces is associated with 
their velocity dependence \cite{Bender:2003}. The expression for $m_{-1}$, in the case of an isovector 
external operator, includes integrals involving ${\cal E}_{\rm sym}$ and $F$. They can be
evaluated in a simple way if one assumes the validity of the leptodermous expansion. 
In this way, one introduces volume and surface coefficients of ${\cal E}_{\rm sym}$ that
can be denoted by $b_{\rm vol}$ and $b_{\rm surf}$, respectively. By specializing 
$F$ to the isovector dipole case, the following expression is obtained:
\begin{equation}\label{Els}
E_{-1}\equiv\sqrt{\frac{m_1}{m_{-1}}}
=\sqrt{\frac{3\hbar^2}{m\langle
r^2\rangle}\frac{b_{\rm vol}}{\left(1+\frac{5}{3}\frac{b_{\rm surf}}
{b_{\rm vol}}A^{-\frac{1}{3}} \right)}(1+\kappa)}.
\end{equation}
This equation has been found to yield values of the centroid energy which
are in rather good agreement with those of microscopic
Hartree-Fock plus Random Phase Approximation (HF-RPA) 
calculations performed with microscopic Skyrme interactions 
\cite{Trippa:2008,Colo:1995}. In fact, the same equation has been used in a previous
study~\cite{Colo:1995}, in order to constrain directly the
parameters of the isovector part of the Skyrme interaction.

Although there is not a straightforward analytic relation between Eq. (\ref{Els}) 
and a similar expression that contains the symmetry energy, 
a semi-analytic relationship has been found in Ref. \cite{Trippa:2008}. The
plausibility of such a relation can be briefly discussed here. 
The coefficient $b_{\rm vol}$ can be identified with $J$. If the nucleus
had a sharp surface this would be the only quantity appearing in
Eq. (\ref{Els}). The relevance of the nuclear surface, however,
manifests itself in the correction $\left( 1+\frac{5}{3}\frac{b_{\rm surf}}
{b_{\rm vol}}A^{-\frac{1}{3}} \right)^{-1}$. One could then assume
that the r.h.s. of Eq. (\ref{Els}) does not scale as 
$\sqrt{b_{\rm vol}} \equiv \sqrt{S(\rho_0)}$, but rather as $\sqrt{S(\bar \rho)}$
where $\bar \rho$ is some value of density below the saturation density 
$\rho_0$, 
namely it is an average density that takes into account the fact that some
nucleons are localised in the inner part of the nucleus where the
density is $\rho_0$ while others are more localised at the surface where
the density is lower. 

\begin{figure}[ht]
\includegraphics[width=0.3\textwidth,angle=-90]{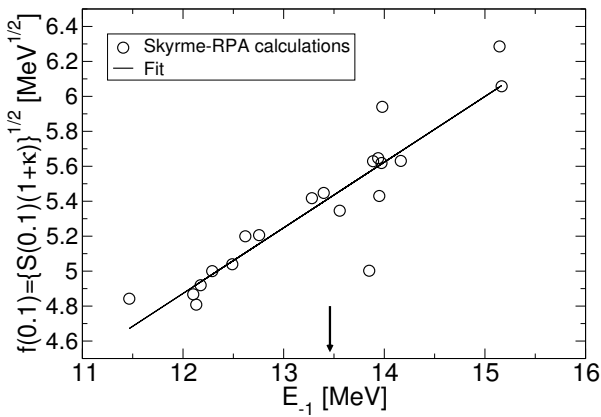}
\caption{Correlation associated with Eq. (\ref{fit1}) in the case of 
$^{208}$Pb. The empty symbols
correspond to the microscopic Skyrme-RPA calculations and the line is a linear
fit whose correlation coefficient is 0.91. The complete information about the
Skyrme models that have been employed, can be found in Ref. \cite{Trippa:2008} 
from which this figure has been adapted. The arrow indicates the experimental
value for $E_{-1}$ from Ref. \cite{Dietrich:1988}.   
\label{fig:s_vs_dipole_energy}}
\end{figure}

In Ref. \cite{Trippa:2008} it has been found indeed that such a correlation
between $E_{-1}$ (calculated within HF-RPA) and $\sqrt{S(\bar \rho)}$
exists. One has to consider the term $(1+\kappa)$, while the value of
$\langle r^2 \rangle$ does not vary significantly when one changes
the model used to calculate the dipole. For heavy nuclei like $^{124}$Sn or $^{208}$Pb the
value of $\bar \rho$ is around 0.1 fm$^{-3}$ whereas this value
tends to lower in, e.g., $^{40}$Ca. The microscopic HF-RPA
results for $E_{-1}$ have been calculated using a broad set of Skyrme 
forces and a strong linear correlation of the type 
\begin{equation}\label{fit1}
\sqrt{S(\bar \rho)(1+\kappa)} = a + b E_{-1}(RPA)
\end{equation}
has been found. Typically, the correlation coefficient is around
$\approx$ 0.9 as it can be seen in Fig. \ref{fig:s_vs_dipole_energy}.
 
Unfortunately, a very precise determination of the
enhancement factor $\kappa$ (in other word, of the total dipole EWSR 
up to high enough energies) is not available. However, by using the case
of $^{208}$Pb and inserting the experimental value for the IVGDR in 
Eq. (\ref{fit1}), it has been found that 
\begin{equation}\label{finalcon}
23.3\;{\rm MeV} < S(0.1) <24.9\;{\rm MeV},
\end{equation}
where the error takes into account the uncertainty of the linear fit
as well as of the experimental value of $\kappa$.

\section{Pygmy Dipole Resonance and other observables related to the dipole
spectra}\label{pdr}

Recently, much interest has been devoted to the dipole strength below
the IVGDR. While in light, neutron-rich halo nuclei this strength may be
specially enhanced due to the large transition probability of weakly bound
neutrons to continuum states (``threshold effect''), this is not the case
in medium-heavy nuclei with neutron excess. These nuclei may be either stable
or unstable but the neutron excess gives rise only to a neutron skin and not
to a halo. In such 
systems a possible, peculiar mode of vibration has been proposed, 
namely the oscillation of the neutrons of the skin with respect to
the (essentially N$\approx$Z) core. The frequency of this mode should be
lower than the IVGDR and the name ``Pygmy Dipole Resonance'' (PDR) has been
introduced in the literature. 

Experimentally, low-lying dipole strength has been found in several
nuclei (see, e.g., Fig. 2 of Ref. \cite{Klimkiewicz:2007}, and the two review papers 
\cite{Paar:2007,Zilges:2013}). To understand the microscopic nature of
this strength, namely to establish whether it corresponds to the PDR
picture that we have just described, or whether it is a non-collective state, 
is hard if not impossible especially when measurements are not exclusive ones.
Typically the PDR strength may arrive up to a few \% of the dipole EWSR.
Among the cases in which this strength appears unambiguously, we mention
the nuclei $^{132}$Sn \cite{Klimkiewicz:2007} and $^{68}$Ni 
\cite{Wieland:2009}. 

\begin{figure*}[ht]
\includegraphics[width=0.6\textwidth]{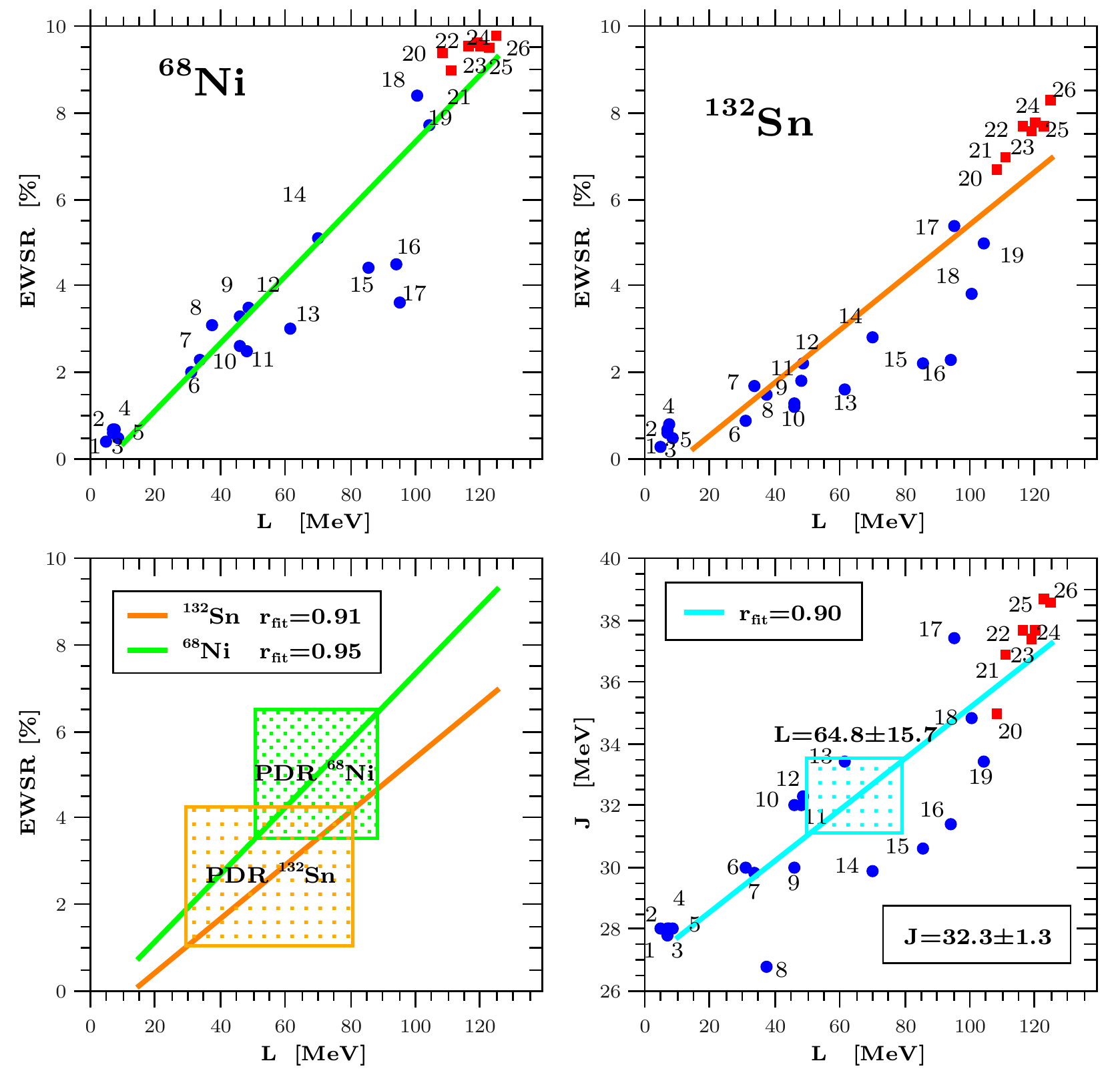}
\caption{In the two upper panels the correlation between the slope parameter
$L$ and the fraction of EWSR exhausted by the PDR is displayed, for the two
nuclei $^{68}$Ni and $^{132}$Sn respectively. In the lower-left panel the results
for the two nuclei are shown together, and the experimental findings are used to
deduce and allowed value for $L$ (as it is discussed in the main text). In the lower-right
panel a correlation plot $L$-$J$ is displayed, so that a value for $J$ is also
deduced. Taken from Ref. \cite{Carbone:2010} (where the detailed correspondence
bwteen numbers and models used can also be found).   
\label{fig:L_vs_PDR_strength}}
\end{figure*}

In Ref. \cite{Carbone:2010}, using those two nuclei, another type of correlation 
between dipole properties and the symmetry energy has been found. 
This correlation is between the percentage of EWSR exhausted by the
PDR and the slope parameter $L$ defined in Eq. (\ref{parameters}). This
correlation appears clearly, although it is not perfect, 
when $L$ is plotted against the fraction
of EWSR calculated not only by means of Skyrme HF-RPA, but by means 
of relativistic RPA on top of Relativistic Mean Field (RMF) as well. By using such 
correlation plot, and the aforementioned experimental data, values of
$L$ have been extracted for the two nuclei under study. 
The separate values of $L$ have been extracted with an error that
takes into account experimental errors and uncertaintes in the fits 
of Fig. \ref{fig:L_vs_PDR_strength}. 
The two values
overlap and if one assumes one can perform a weighted average 
for both mean values and uncertainties, one obtains 
\begin{equation}
L = 64.8 \pm 15.7\ \rm MeV.
\label{l-pdr}
\end{equation}  
Within the choice of our models, this implies $J$ = 32.3$\pm$1.3 MeV.
All these points are illustrated in Fig. \ref{fig:L_vs_PDR_strength}.

In Ref. \cite{Carbone:2010}, a careful analysis has also been made
concerning the neutron skins. As it has been demonstrated by more than 
one author \cite{Brown:2000,Furnstahl:2002}, the neutron skin is in fact
well correlated with $L$ (this can be understood, since if $L$ increases 
the symmetry energy undergoes larger changes going from the
inner part of the nucleus to the surface, and so the system finds it energetically
more convenient to localise the excess neutrons on the surface). 
Consequently, the values of the neutron skins $\Delta R$ extracted from the
previous value of $L$ are
\begin{eqnarray}
\Delta R \left( ^{68}{\rm Ni} \right) & = & 0.200\pm0.015\ {\rm fm}; \nonumber \\ 
\Delta R \left( ^{132}{\rm Sn} \right) & = & 0.258\pm0.024\ {\rm fm}; \nonumber \\ 
\Delta R \left( ^{208}{\rm Pb} \right) & = & 0.194\pm0.024\ {\rm fm}. 
\label{dr-pdr}
\end{eqnarray}  

We do not dispose of a model that can help as a guideline 
to understand in detail the correlation between the EWSR of the PDR 
and $L$. In a simplified picture, if the PDR is really a mode in which
the excess neutrons participate and their dynamics is decoupled from the 
IVGDR, then the correlation we have discussed can be intuitively 
understood. In fact, the percentage of EWSR of the PDR increases with
the number of neutrons belonging to the ``skin'' \cite{Suzuki:1990}, and
the skin increases with $L$ as we have just discussed. 

However, this picture may not be valid in all nuclei where the PDR shows up.
In Ref. \cite{Nazarewicz:2010}, it has been claimed that the PDR has a
single-particle character: this analysis has been carried out using a specific
local energy functional, namely SV-bas. In Ref. \cite{Roca-Maza:2012a} the
microscopic character of the PDR has been analysed in detail in the 
nuclei $^{68}$Ni, $^{132}$Sn and $^{208}$Pb: while for certain Skyrme energy
functionals (those characterised by a larger value of $L$) the PDR
shows up as a well-defined peak so that it can be truly defined as a resonance,
in other cases this does not happen. Consequently, it may be more cautious
to refer to the low-lying dipole strength as composed of ``pygmy dipole
states'' (PDSs). These states have also a mixed isospin character and in the
isoscalar response they show up more clearly, and display more coherence
of the microscopic particle-hole (p-h) amplitudes, than in the isovector
response. A similar analysis, leading to consistent conclusions, has been
performed in Ref. \cite{Vretenar:2012}. A possible component of toroidal
motion in the low-energy dipole strength had been initially claimed in the 
microscopic calculations of Ref. \cite{Vretenar:2001} and recently
revived (see, e.g., \cite{Repko:2013} and references therein). About the nature
of the pygmy states, and the question whether any relation with the symmetry
energy or the neutron skin has to be expected, the reader can also consult
the contribution by J. Piekarewicz to this volume (and references therein), 
as well as the works on the PDR by T. Inakura and co-workers 
\cite{Inakura:2011,Inakura:2012}.

In conclusion, the nature of the PDR is still strongly debated because of
its complex and mixed (isoscalar/isovector, surface/volume, irrotational/toroidal)
character. Certainly, more experimental and theoretical investigation should
be envisaged. For the time being, it remains intriguing how despite all
warnings the EWSR carried by the PDR could lead to very reasonable values
of $J$ and $L$ as it has been shown in the first part of this Section.

We do not discuss in this contribution other aspects related to IV dipole
observables and the symmetry energy. The relevance of the total
dipole polarizability \cite{Piekarewicz:2012}, and the fact that $L$ turns out
to be correlated with the product of $J$ times the dipole polarizability 
\cite{Roca-Maza:2013b}, are thoroughly discussed in the contribution to
this volume by J. Piekarewicz.

\section{Symmetry energy from the IVGQR}\label{ivgqr}

The properties of the isovector giant quadrupole resonance (IVGQR) 
have not been determined so accurately for some time, 
due to the lack of experimental probes having good selectivity. Recently, at the
HI$\vec\gamma$S facility, it has been demonstrated that
the scattering of polarized photons can provide a direct measurement
of the IVGQR properties, without the uncertainties associated with
hadronic probe experiments \cite{Henshaw:2011}. This experimental
achievement has motivated the attempt to extract information about the symmetry
energy. 

The energy of the IVGQR receives contribution from the unperturbed p-h configurations
at  2$\hbar\omega$ excitation energy, plus some correlation energy related
to the residual interaction. Since the residual interaction is in the isovector
channel, it can be naturally linked with the symmetry energy. In Ref.
\cite{Roca-Maza:2013a} the quantum harmonic oscillator model has
been applied to the IVGQR case and, with mild assumptions and taking care
of the fact that the unperturbed energy can be related to the effective mass
and, in turn, to the isoscalar GQR energy, the main result is the following formula:
\begin{equation}
E_{\rm IVGQR} \approx 2 \left[\frac{\left(E_{\rm ISGQR}\right)^2}{2}+ 
2\frac{\varepsilon_{{\rm F}_\infty}^2}{A^{2/3}}\left(\frac{3S(\rho_A)}{\varepsilon_{{\rm F}_\infty}} - 
1\right)\right]^{1/2},
\label{ex-ivgqr-3}
\end{equation}
where $\varepsilon_{{\rm F}_\infty}$ is is the Fermi energy for
symmetric nuclear matter at saturation density, and $S(\rho_A)$ is the symmetry energy
at some average nuclear density for the nucleus having mass number A. As we have done
above in the IVGDR case, we can choose this average density as 0.1 fm$^{-3}$. A first
important outcome of the previous equation is that the same value of $S(0.1)$ that
we have derived in Eq. (\ref{finalcon}) is consistent with the experimental energies 
of the ISGQR and IVGQR, and turns out to be further validated by this fact. 

\begin{figure}[ht]
\includegraphics[width=0.45\textwidth]{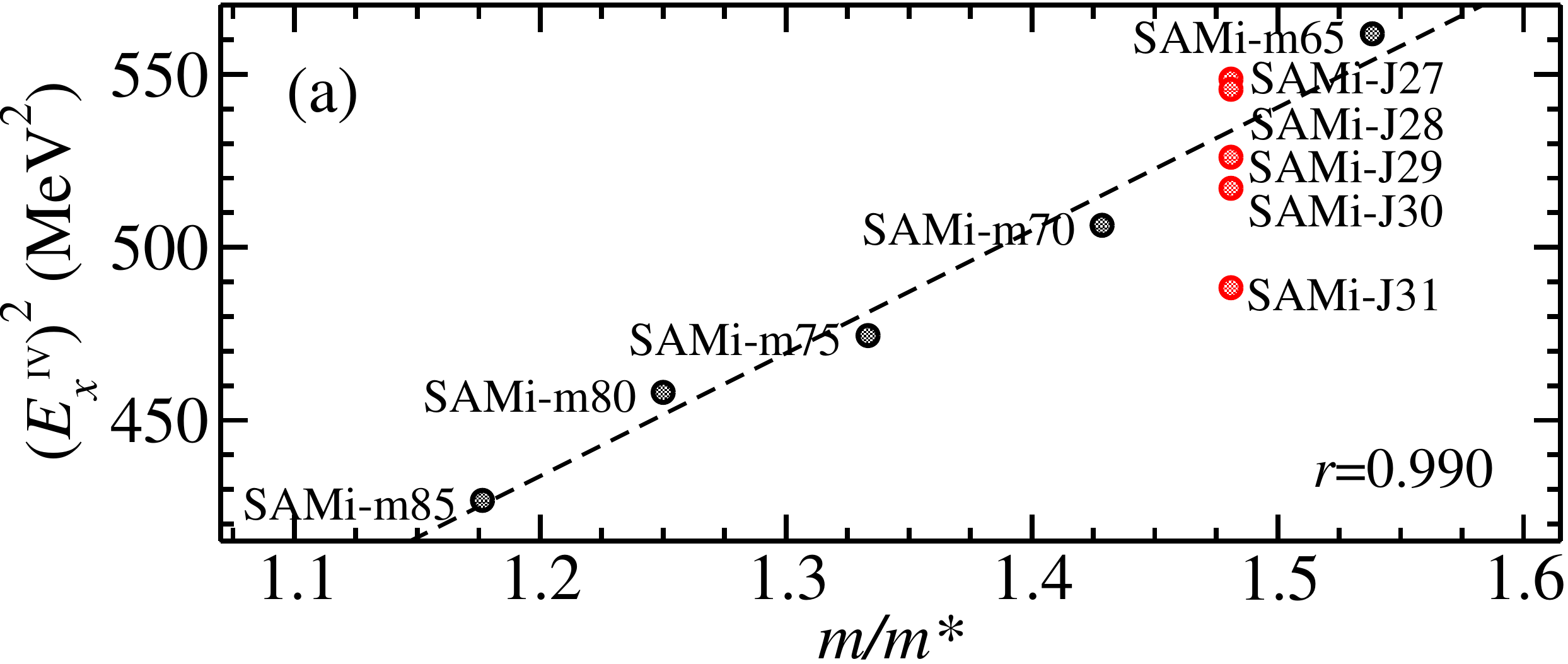}\\
\includegraphics[width=0.45\textwidth]{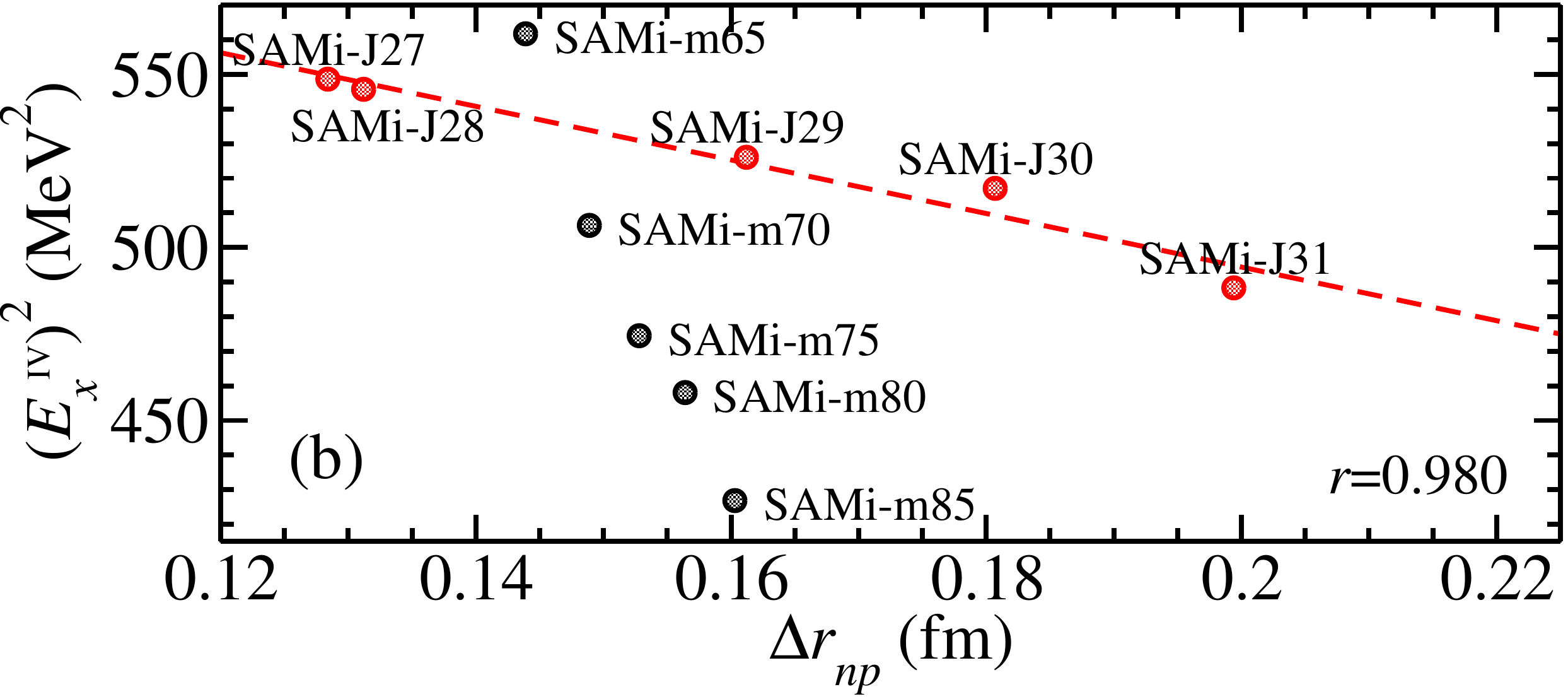}
\caption{Sensitivity of the energy of the IVGQR in $^{208}$Pb to the effective mass (upper panel)
and to the neutron skin (lower panel). For details about the models whose names appear in
the panels, and for an explanation of the emerging correlations, see the main text. Taken from
Ref. \cite{Roca-Maza:2013a}.
\label{fig:gqr}}
\end{figure}

Both Skyrme and relativistic mean-field models do indeed follow quite well the
scaling predicted by Eq. (\ref{ex-ivgqr-3}). In the case of the Skyrme models, we demonstrate
this fact in Fig. \ref{fig:gqr}. The models have been built by using the same protocol of the recent
parameter set SAMi \cite{Roca-Maza:2012b}. In some of them all nuclear matter parameters
have been kept fixed as in the original force, but the effective mass has been changed: thus,
SAMi-m85 means that $m*/m$ is 0.85; in others, all nuclear matter properties have been 
kept fixed but the symmetry energy at saturation has been changed: in this case, SAMi-J27
means that $J$=27 MeV. 

From the upper panel of Fig. \ref{fig:gqr} it is clear that if the
symmetry energy properties are kept fixed, the energy increases if the square root of the
effective mass decreases. This is due to the first term of Eq. (\ref{ex-ivgqr-3}), as the
ISGQR energy is known to scale with $\sqrt{m/m^*}$ \cite{Blaizot80}. 
In the same way, it is expected that the second term in Eq. (\ref{ex-ivgqr-3}) plays a 
role and produces an increase of the IVGQR energy with $S(0.1)$. Actually
in our fitting protocol we fix $J$, and for increasing values of this quantity the result of the
fit is a comparatively more important decrease at the same time of $L$ and $S(0.1)$:
therefore, both the neutron skin and the IVGQR energy decrease, displaying consequently the
interesting correlation of the lower panel of Fig. \ref{fig:gqr}. If we use such
kind of correlations between the IVGQR energy and the properties of the symmetry energy,
assuming a value of $J$ = 32$\pm$1 MeV we extract
\begin{eqnarray}
L & = & 37 \pm 18\ {\rm MeV}; \label{l-gqr}\\
\Delta R \left( ^{208}{\rm Pb} \right) & = & 0.14\pm0.03\ {\rm fm}. \label{dr-gqr} 
\end{eqnarray} 
Note that these values are lower but still compatible with the results (\ref{l-pdr}) and (\ref{dr-pdr}) 
extracted from the PDR.

\section{Nuclear Incompressibility and symmetry energy parameters}\label{incompr}
The  incompressibility $K_{\infty}$ of infinite symmetric matter  is defined by the second 
 derivative of the energy per particle ${\cal E}_0(\rho,\beta=0)/\rho$ with respect to the density 
$\rho$ at the saturation point,  
\begin{equation}
 K_{\infty}=9\rho^2\frac{d^2}{d\rho^2} \left. \left(\frac{ {\cal E}_0(\rho,\beta=0)}{\rho} \right) 
\right|_{\rho=\rho_{0}},
\label{eq:Knm}
\end{equation}
where $ {\cal E}_0(\rho,\beta=0)$ is the isoscalar part of the energy density $ {\cal E}(\rho,\beta)$ for nuclear 
matter given in Eq. (3).
The nuclear matter incompressibility $K_{\infty}$ is not a directly 
measurable quantity. Instead, the energy of isoscalar giant monopole resonance,  $E_{ISGMR}$, is expressed 
in terms of the finite nucleus incompressibility $K_A$ as  \cite{Strin82,trein81}
\begin{equation}
E_{ISGMR}=\sqrt{\frac{\hbar^2K_A}{m<r^2>_m}},
\label{eq:E_ISGMR}
\end{equation}
where $m$ is the nucleon mass and $<r^2>_m$ is the mean square 
mass radius of the ground state.  The finite nucleus incompressibility 
can be parameterized by means of a similar expansion 
as the liquid drop mass formula with the volume, surface, symmetry and Coulomb 
 terms \cite{Blaizot80}:
\begin{equation}
K_A=K_{vol}+K_{surf}A^{-1/3}+K_{\tau}\delta^2+K_{Coul}\frac{Z^2}{A^{4/3}},
\label{eq:K_A}
\end{equation}
where $\delta=(N-Z)/A$. We denote by $K_{\tau}$ the asymmetry term of the finite nucleus 
 incompressibility 
$K_{A}$  because the symbol $K_{sym}$ has been already used 
as one of the isovector nuclear matter properties in Eq. (5).  
The volume 
term
 $K_{vol}$  
of the finite nucleus incompressibility $K_{A}$  is 
directly related to 
the nuclear matter incompressibility $K_{\infty}$.   
The asymmetry term  $K_{\tau}$ is related to nuclear matter properties 
as \cite{Blaizot80,Sagawa2007} 
\begin{equation}
 K_{\tau}=K_{sym}+3 L -L B,
\label{Ka-sym-eq}
\end{equation}
where $B$ is proportional to the third derivative of the energy
density with respect to the density at the saturation point,
\begin{equation}
  B=\frac{27 \rho_{0}^{2}}{K_{\infty}} \left.
\frac{d^{3} {{\cal E}_0(\rho,\beta=0)}}{d \rho^{3}} \right|_{\rho=\rho_{0}}.
\label{eq:B}
\end{equation}
In Refs. \cite{Chen2009,cdplb},  $K_{\tau}$ is expressed in terms of the skewness parameter $Q$ as
\begin{equation}
 K_{\tau}=K_{sym}-6 L -\frac{L Q}{K_{\infty}}, 
\label{Ka-sym-eq2}
\end{equation}
where $Q$ is defined as the third derivative of the energy per nucleon with respect to the density at the saturation point,
\begin{equation}
  Q=27 \rho_{0}^{2} \left.
\frac{d^{3} ({\cal E}_0(\rho,\beta=0)/\rho)}{d \rho^{3}} \right|_{\rho=\rho_{0}}.
\label{eq:Q}
\end{equation}
The formulas (\ref{Ka-sym-eq}) and (\ref{Ka-sym-eq2}) look different, but they are equivalent since $B$ and $Q$ are related by the equation
\begin{equation}
 B=9+\frac{Q}{K_{\infty}}.
 \end{equation}
 
The analytic formulas for $K_{surf}$ and $K_{Coul}$ are given by
\begin{eqnarray}
K_{surf}=4\pi{r_0}^2 \left[ 4 \sigma(\rho_{0}) +9 \rho_{0} \left. 
\frac{d^2 \sigma}{d \rho^2} \right|_{\rho=\rho_{0}} 
+2\sigma(\rho_{0})B
 \right],\nonumber \\
\label{Ksurf}
\end{eqnarray}
\begin{eqnarray}
K_{Coul}=\frac{3}{5}\frac{e^2}{r_0}(1-B),
\label{Kcoul}
\end{eqnarray}
where $r_{0}$ is the radius constant defined by
\begin{equation}
r_{0}= \left(\frac{3}{4\pi\rho_{0}}\right)^{1/3}.
\end{equation}
In Eq. (\ref{Ksurf}), $\sigma$ is the surface tension in symmetric semi-infinite nuclear matter 
defined by
\begin{equation}
\sigma(\rho_{0})=\int_{-\infty}^{\infty} \left[ {\cal E}(\rho,\beta=0)-
\frac{{\cal E}_0(\rho_0,\beta=0)}{\rho_{0}} \rho \right] 
dz.  
\end{equation}
$K_{surf}$ can be evaluated by the extended Thomas-Fermi approximation and 
the scaled HF calculations of semi-infinite nuclear matter in the Skyrme Hartree-Fock (SHF) model.
These evaluations show that the approximate relation $K_{surf}\sim -K_{\infty}$ holds    
within an accuracy of a few \% in the SHF model.  In relativistic models, the extended Thomas-Fermi 
approximation gives as a result a slightly larger surface contribution, 
for example,  $K_{surf}\sim -1.16K_{\infty}$ in the case of NL3.

\begin{figure}[h]
\includegraphics[width=0.5\textwidth]{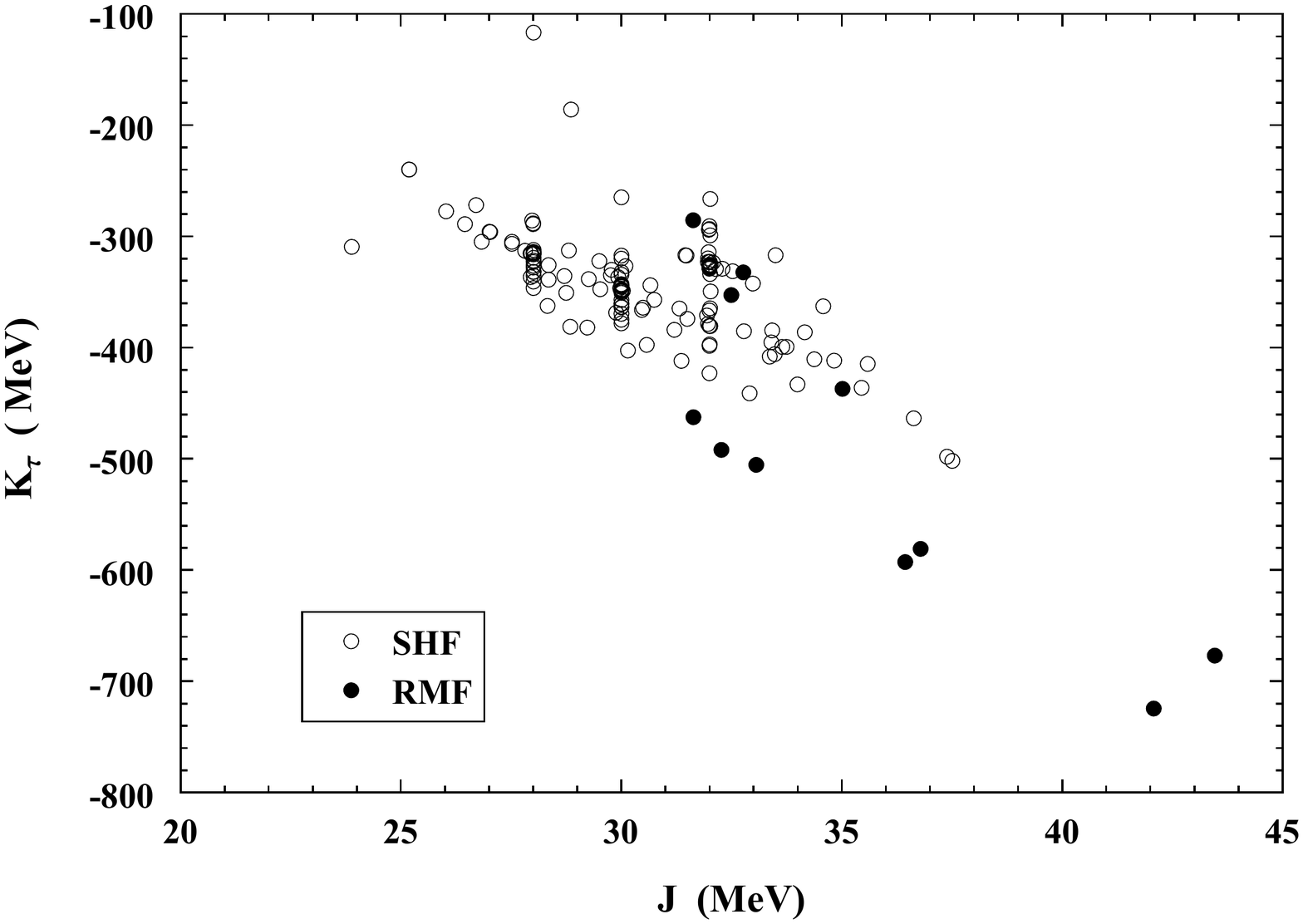}
\caption{Correlation between  the asymmetry term of the finite nucleus incompressibility $K_{\tau}$  and  the volume symmetry  energy  $J$, calculated by using various 
Skyrme parameter sets (SHF, open circles) and relativistic Lagrangians (RMF, filled circles).  
\label{fig:Ktau-J}}
\end{figure}

\begin{figure}[h]
\includegraphics[width=0.5\textwidth]{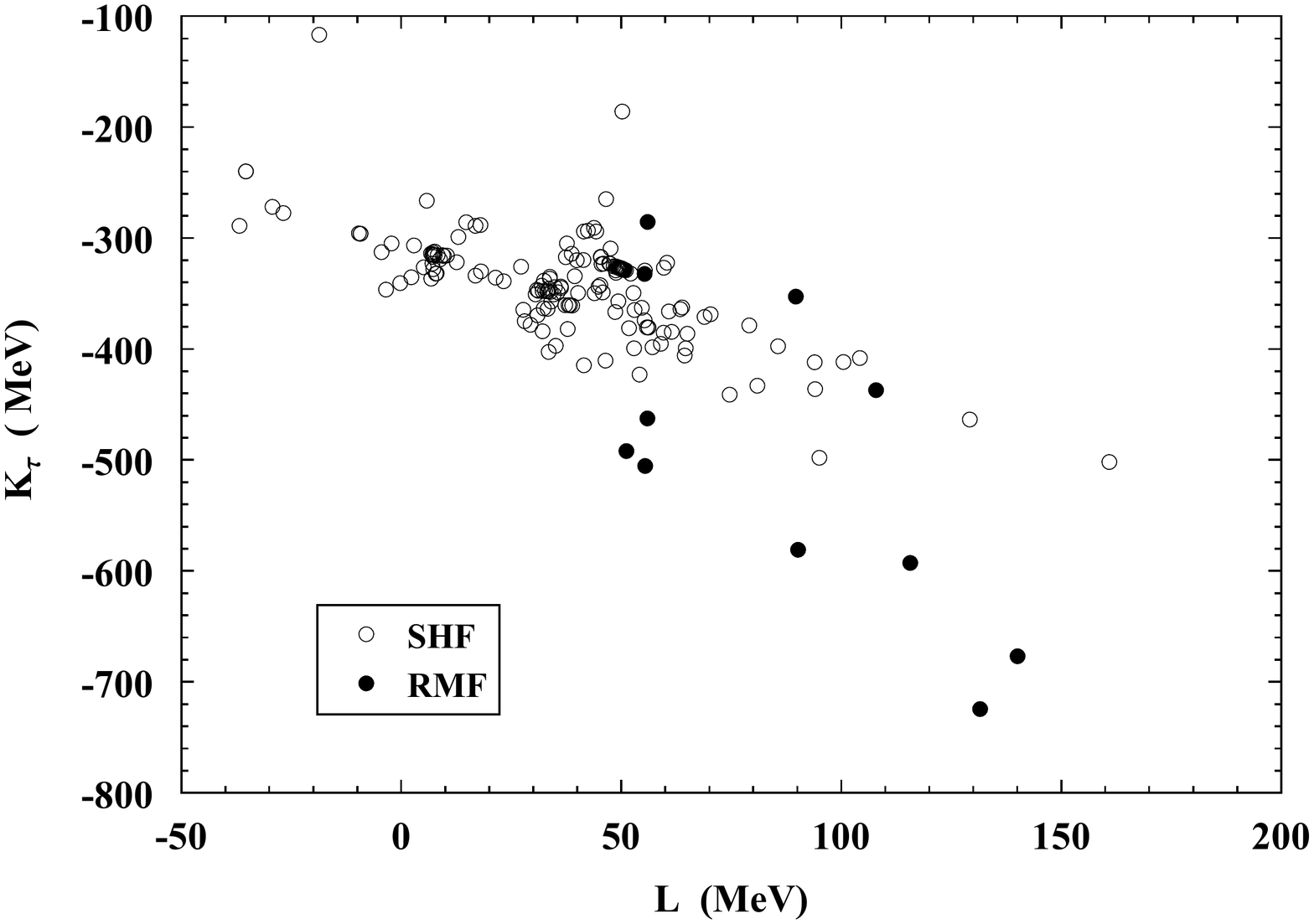}
\caption{Correlation between  the asymmetry term of the finite nucleus incompressibility $K_{\tau}$  and  the slope parameter $L$,  calculated by using various 
Skyrme parameter sets (SHF, open circles) and relativistic Lagrangians (RMF, filled circles).    
\label{fig:Ktau-L}}
\end{figure}

\begin{figure}[h]
\includegraphics[width=0.5\textwidth]{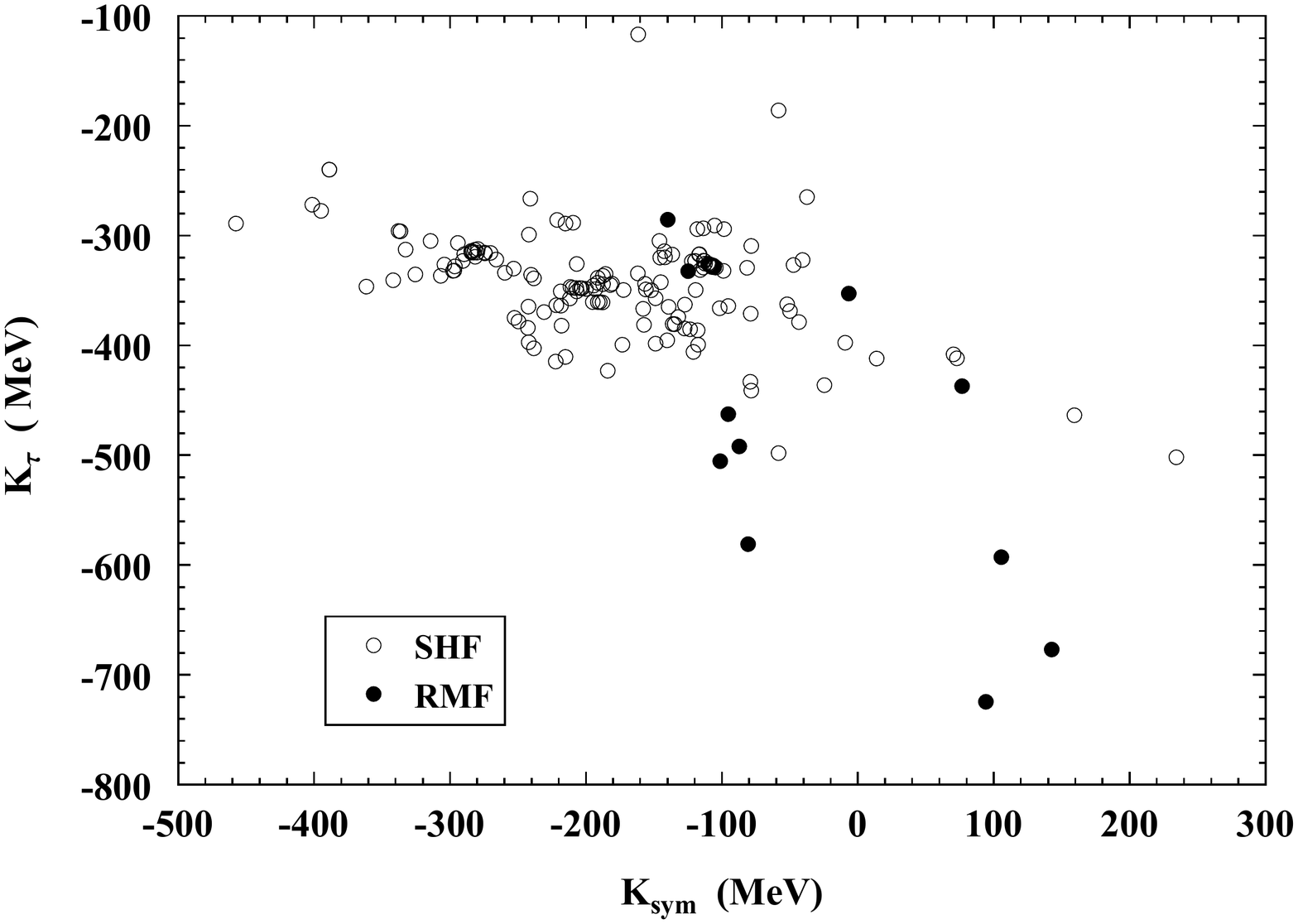}
\caption{Correlation between  the asymmetry term of the finite nucleus incompressibility $K_{\tau}$  and  the second derivative of the 
symmetry  energy  $K_{sym}$,  calculated by using various 
Skyrme parameter sets (SHF, open circles) and relativistic Lagrangians (RMF, filled circles).  
\label{fig:Ktau-Ksym}}
\end{figure}

It is feasible to calculate the values of $K_{\tau}$ and  $K_{Coul}$  by using
various Skyrme Hamiltonians and relativistic Lagrangians.  
 It was pointed out in Ref. \cite{Yoshida2006} that
 there are no clear correlations between $L$ and  $K_{\infty}$,   and 
 between   $K_{sym}$ and  $K_{\infty}$.  On the other hand it is remarkable
 that $K_{\tau}$, as a linear combination of  $K_{sym}, L$ and  $B$, 
 show a clear correlation with  $K_{\infty}$ having a large correlation
 coefficient, especially in SHF models  \cite{Yoshida2006}.  
Correlation plots between $K_{\tau}$, the symmetry coefficient of the nuclear incompressibility, 
and the parameters  
$J$, $L$ and $K_{sym}$ characterising the symmetry energy, are displayed in Figs. \ref{fig:Ktau-J},  \ref{fig:Ktau-L} and \ref{fig:Ktau-Ksym}, respectively \cite{Yoshida2013}.  
The  plot involving $J$ and $K_{\tau}$ in Fig. \ref{fig:Ktau-J} shows a clear 
 correlation with a negative slope. We can see also a similar correlation 
between 
  $L$  and  $K_{\tau}$ in  Fig. \ref{fig:Ktau-L}. Both correlation 
coefficents are between -0.6 and -0.7. Thus, empirical information on   $K_{\tau}$ 
from e.g. the isotopic dependence of the ISGMR energies may  give some 
constraint on these two values.  We will come back to this point at the
end of Sec. 6. On the other hand, the points in the plot of 
 $K_{sym}$ vs.  $K_{\tau}$ are rather scattered, and it looks difficult to constrain the value of
 $K_{sym}$ from the empirical value of $K_{\tau}$.  
 We should also  make the remark
 that the results of RMF and 
 Skyrme models have rather large overlap and give consistent  results in 
 the correlation plots of Figs. \ref{fig:Ktau-J}  and \ref{fig:Ktau-L}.

It was pointed out that  any Hamiltonian 
which has a larger $K_{\infty}$  gives a smaller $K_{\tau}$ \cite{Sagawa2007}.
The variations  of   $K_{\tau}$ for the Skyrme interactions are
\begin{equation}
K_{\tau} =(-400\pm100) {\rm ~MeV}   \hspace{1cm} \mbox{for Skyrme interactions}.
\end{equation}
On the other hand, the values of RMF are largely negative and 
have  more variation among the seven effective relativistic mean field (RMF) Lagrangians,
\begin{equation}
K_{\tau} =(-620\pm180) {\rm ~MeV}   \hspace{1cm} \mbox{for RMF Lagrangians}.
\end{equation}
 In principle,
the value of $K_{Coul}$ should be model-in\-de\-pen\-dent.  
Among the 13 
parameter sets of Skyrme interactions, the variation of $K_{Coul}$ 
 is rather small,
\begin{equation}
  K_{Coul} =(-5.2\pm0.7) {\rm ~MeV}   \nonumber
\end{equation}
compared with that of $K_{\tau}$.  The values of $K_{Coul}$ in RMF show
essentially the same trend, but have a larger variation.

\section{Nuclear incompressibility and the asymmetry term from the ISGMR}\label{isgmr}

The study of the isoscalar giant monopole resonance (ISGMR) provides a direct experimental connection to nuclear incompressibility in finite nuclear systems. The centroid energy of ISGMR, $E_{ISGMR}$, can be related to the nuclear incompressibility of finite nuclear matter, $K_A$, as given by Eq. (\ref{eq:E_ISGMR}).
The ISGMR strength distribution can be determined experimentally via inelastic scattering of isoscalar probes. The most commonly-used, and effective, probe for such investigations has been the $\alpha$-particle (the $^4$He nucleus). In these investigations, inelastic scattering measurements are performed off a particular target at very forward angles, including 0$^{\circ}$.

The importance of making measurements at such extreme forward angles, including $0^\circ$, is twofold: the cross section for the ISGMR peaks at $0^\circ$, and the $L$=0 angular distribution is most distinct at the very forward angles. These measurements are, however, extremely difficult since the primary beam passes very close to the scattered particles at these angles and one requires a combination of a high-quality, halo-free beam, and an appropriate magnetic spectrometer. The high-resolution magnetic spectrometer Grand Raiden, at the Research Center for Nuclear Physics (RCNP) at Osaka University, Japan \cite{Fujiwara99} is a most suitable such instrument; similar measurements are being carried out at the 
Texas A \& M University cyclotron facility as well \cite{dhyxx}. An unmatched asset of Grand Raiden is that its 
optical properties allow for collection of inelastic scattering spectra practically free of all instrumental background that had been a bane of such measurements in the past. 

The inelastic scattering spectra are analyzed using multipole decomposition analysis (MDA) \cite{Bonin84,Itoh2003} to extract the ISGMR strength distributions, the centroid of which can give the compressibility, $K_A$ of the nucleus under investigation. 
Examples of such ``background-free''' spectra, as well as the details of the experimental techniques and analysis procedures for these measurements have been provided in several recent reports from the RCNP work (see, for example, Refs. \cite{Tao2007,Tao2010,cdplb}).

To go from $K_A$ to $K_{\infty}$, one
builds a class of energy functionals, $E(\rho)$ [cf. Eq. (\ref{edf})], with different parameters that allow calculations for nuclear matter and finite nuclei in the same theoretical framework. The parameter-set for a given class of energy functionals is characterized by a specific value of $K_{\infty}$. The ISGMR strength distributions are obtained for different
energy functionals in a self-consistent RPA calculation. The $K_{\infty}$ associated with the interaction that best reproduces the ISGMR centroid energies is, then, considered the correct value \cite{Blaizot80}.

Following this procedure, both relativistic and non-relativistic calculations give $K_{\infty}$=240 $ \pm $ 20 MeV \cite{shlomo,colo2004,jorge1,shlomo:2006,colo2008}.
These accurately calibrated relativistic and non-relativistic models reproduce very well the ISGMR centroid energies in the ``standard" nuclei, $^{90}$Zr and $^{208}$Pb. However, it has been established in recent measurements on the Sn and Cd isotopes \cite{Tao2010,cdplb} that this value of $K_{\infty}$ significantly overestimates $E_{ISGMR}$ for these ``open shell'' nuclei. In other words, it would appear that the Sn and Cd nuclei are ``softer'', considering the $E_{ISGMR}$ from just these nuclei would yield an appreciably lower value for $K_{\infty}$.
Pairing correlations have been suggested as a reason for this softening; yet, the results are not conclusive 
\cite{li,vesely,cao}. 

As noted in the previous Section, $K_A$ may be parameterized as:
\begin{eqnarray}\label{eq:3}
K_A\approx K_{vol}(1+cA^{-1/3})+K_\tau\left( \frac{N-Z}{A} 
\right)^2+K_{Coul}\frac{Z^2}{A^{4/3}}.
\nonumber \\
\end{eqnarray}
\noindent
Here, 
 $c\approx -1$ as noted previously and discussed in detail in Ref. \cite{Patra2002}; $K_{Coul}$ is essentially a model-independent term (in the sense that the deviations from one theoretical model to another are quite small) \cite{Sagawa2007}; and $K_{\tau}$ is the asymmetry term.
Although closely related, the finite-nucleus asymmetry 
term $K_{\tau}$ should not be confused with the corresponding 
term in infinite nuclear matter--a quantity also denoted 
by $K_{\tau}$ at times, but which should actually be written as
$K_{\tau}^{\infty}$ 
(we have introduced this quantity in Eq. (18) above, and showed
that it should not be confused with $K_{sym}$ either; in fact, 
the asymmetry coefficient of the finite nucleus incompressibility
does not take contribution merely from the second derivative of the
symmetry energy).
$K_{\tau}^{\infty}$ should never 
be regarded as the $A\!\rightarrow\!\infty$ limit of the finite-nucleus 
asymmetry $K_{\tau}$.  
Yet the fact that $K_{\tau}$ is both experimentally accessible and strongly correlated with $K_{\tau}^{\infty}$ is vital in placing stringent constraints on the
density dependence of the symmetry energy. 
It is the strong sensitivity of $K_{\tau}^{\infty}$ to the density dependence of the symmetry energy that makes this investigation of critical importance in constraining the EOS of neutron-rich matter. 

This asymmetry term,  $K_{\tau}$, can be obtained by investigating the ISGMR over a series of isotopes for which the neutron-proton asymmetry, $(N-Z)/A$, changes by an appreciable amount. Coming back to Eq. (\ref{eq:3}), for a series of isotopes, the difference $K_A-K_{Coul}Z^2A^{-4/3}$ may be approximated to have a quadratic relationship with the asymmetry parameter ((N - Z)/A)), of the type y = A + Bx$^2$, with $K_\tau$ being the coefficient, B, of the quadratic term. 

\begin{figure}[h]
\includegraphics[width=0.5\textwidth]{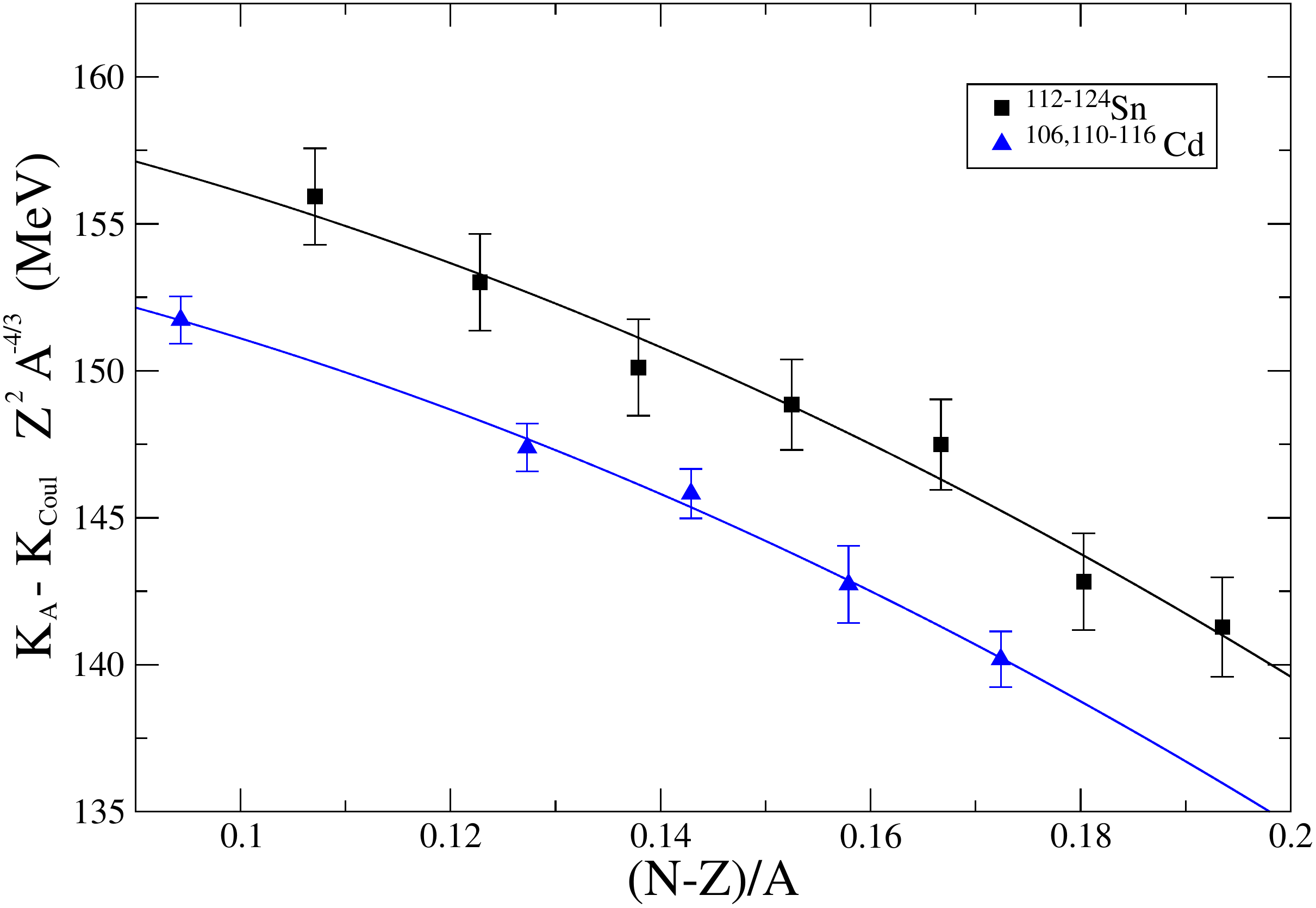}
\caption{The difference $K_A-K_{Coul}Z^{2}A^{-4/3}$ in the Sn and Cd isotopes plotted as a function of the asymmetry parameter, $(N-Z)/A$. The data are from Refs. \cite{Tao2010,cdplb}. The values of $K_A$ have been derived using the customary moment ratio $\sqrt{m_1/m_{-1}}$ for
the energy of ISGMR, and a value of 5.2 $\pm$ 0.7 MeV has been used for $K_{Coul}$ (see previous Section). The solid lines correspond to $K_{\tau}$ = - 550 MeV.
\label{fig:ktau}}
\end{figure}
\begin{figure}[h]
\includegraphics[width=0.5\textwidth]{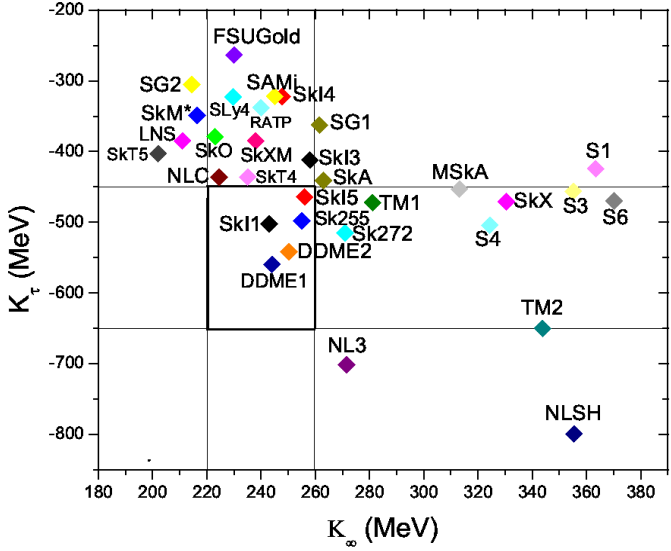}
\caption{Values of K$_\infty$ and K$_\tau$ calculated from the parameter sets of various interactions as labeled \cite{Sagawa2007}.
The vertical and horizontal lines indicate the experimental ranges of K$_\infty$ and K$_\tau$, as determined from the GMR work.
\label{fig:kinftyktau}}
\end{figure}

Such an investigation was carried out by Li {\it et al.} over the even-$A$ $^{112-124}$Sn isotopes \cite{Tao2007,Tao2010} and by Patel {\it et al.} over the even-$A$ $^{106,110-116}$Cd isotopes \cite{cdplb}. The Sn isotopes yielded a value of $K_{\tau} = -550 \pm$100 MeV, the Cd isotopes resulted in $K_{\tau} = -555 \pm$75 MeV. Not only are the two values thus obtained in excellent agreement with each other, but also are consistent with values indirectly obtained from several other measurements: $K_\tau=-370 \pm$120 MeV obtained from the analysis of the isotopic transport ratios in medium-energy heavy-ion reactions \cite{Chen2009}, $K_\tau=-500^{+120}_{-100}$ MeV obtained from constraints placed by neutron-skin data from anti-protonic atoms across the mass table \cite{cente09}; and, $K_\tau=-500 \pm$50 MeV obtained from theoretical calculations using different Skyrme interactions and relativistic mean-field (RMF) Lagrangians \cite{Sagawa2007}. 
In Fig.~\ref{fig:ktau}, we show the data for the Sn and Cd isotopes from Refs. \cite{Tao2010,cdplb} along with quadratic fits with a common value of $K_{\tau} = - $550 MeV.

From the correlation plots in Figs. \ref{fig:Ktau-J} and
\ref{fig:Ktau-L}, one may extract the symmetry energy
coefficients $J$ and $L$ from the empirical value
$K_{\tau} = - $550 MeV. We must take into account the error
on this latter quantity ($\pm$ 100 MeV), as well as
the uncertainties on the linear fits. In this way,
$J$ is found to lie in the range 27.7-35.6 MeV.
On the other hand, the correlation between $K_\tau$ and
$L$ is weaker and we cannot get a meaningful
constraint on $L$.

The ``experimental'' values thus obtained from the ISGMR for $K_{\infty}$ and $K_{\tau}$ taken together can provide a means of selecting the most appropriate of the interactions used in EOS calculations. In Fig.~\ref{fig:kinftyktau}, we plot the $K_{\infty}$ and $K_{\tau}$ for a number of interactions used in nuclear structure and EOS calculations. 
It would appear, indeed, that a vast majority of the interactions fail to meet the criterion established by these measurements. 
A caveat to this statement, though: the $K_{\tau}$ obtained in these measurements is only an ``average'' value, and the data cannot disentangle the volume 
symmetry from higher-order effects like the surface symmetry. Thus, this average value has been identified with the volume symmetry only,
and compared with the volume symmetry coefficient provided by the models.
It is possible, then, to execute similar fits including higher-order terms and obtain very different values for $K_{\tau}$ \cite{pearson}; however, the ``appropriateness'' of the values of the extra terms thus obtained remains unclear. 

\section{Spin-dipole resonances and neutron skin}\label{sdr}
As was mentioned in Section 3, the neutron skin gives an important information about the constraints on the symmetry energy.    It is known that  the model-independent 
non energy-weighted sum rule of charge exchange 
spin-dipole (SD) excitations is directly related to 
the neutron skin thickness \cite{Gaarde}. 
Recently, SD excitations were studied in $^{90}$Zr by the charge-exchange reactions 
 $^{90}$Zr(p,n)$^{90}$Nb \cite{Wakasa} and  $^{90}$Zr(n,p)$^{90}$Y 
 \cite{Yako}, and the model-independent sum rule for the
SD excitations were extracted in Ref. \cite{Yako06}  by using multipole
decomposition analysis (MDA)  \cite{Ichi}.
 The charge exchange reactions 
($^3$He,$t$) on Sn isotopes
were also studied to extract the neutron skin thickness 
 \cite{SD-Pb}.
However, one needs the counter experiment ($t, ^3$He) or (n,p)
on Sn isotopes 
in order to extract
the model-independent sum rule value from experimental data. This counter 
experiment is missing in the case of Sn isotopes.

  The operators for   $\lambda -$pole SD transitions are defined as 
\begin{eqnarray}
\hat{S}^{\lambda}_{\pm } =
 \sum_{i} t_{\pm}^{i}r_{i}[\sigma \otimes Y_{l=1}(\hat{r}_{i})]^{\lambda=0,1,2}, 
\label{eq:eq1}
\end{eqnarray} 
 with the isospin operators being denoted as 
 $t_{\pm} =  t_{x}\pm it_{y}$. 
The model-independent sum rule for the
  $\lambda -$pole SD operator $\hat{S}^{\lambda}_{\pm } $
 can be 
obtained as 
\begin{eqnarray}
& S^{\lambda}_{-}&-S^{\lambda}_{+}=
 \sum _{i \in all} \mid \langle i\mid\ \hat{S}^{\lambda}_{-} \mid0\rangle
 \mid ^2 -
\sum _{i \in all} \mid \langle i\mid\ \hat{S}^{\lambda}_{+} \mid0\rangle
 \mid ^2      \nonumber \\
&=& \langle 0\mid\ [\hat{S}^{\lambda}_{-}, \hat{S}^{\lambda}_{+}] \mid0\rangle
 = \frac{(2\lambda+1)}{4\pi}(N\langle r^2\rangle _n -Z\langle r^2\rangle _p).
 \label{eq:sum_sd_a}
\end{eqnarray}
 The sum rule for the spin-dipole operator (\ref{eq:eq1})
then becomes
\begin{eqnarray}
 S_{-}- S_{+}=\sum_{\lambda}(S^{\lambda}_{-}- S^{\lambda}_{+})=
\frac{9}{4\pi}(N\langle r^2\rangle _n -Z\langle r^2\rangle _p).
 \label{eq:sum_sd}
\end{eqnarray}
It should be noted that 
the sum rule (\ref{eq:sum_sd}) is directly related to the difference between
the mean square radius of neutrons and protons 
with the weight of neutron and 
proton numbers.  

\begin{figure}[htp]
\includegraphics[width=3.8in,clip]{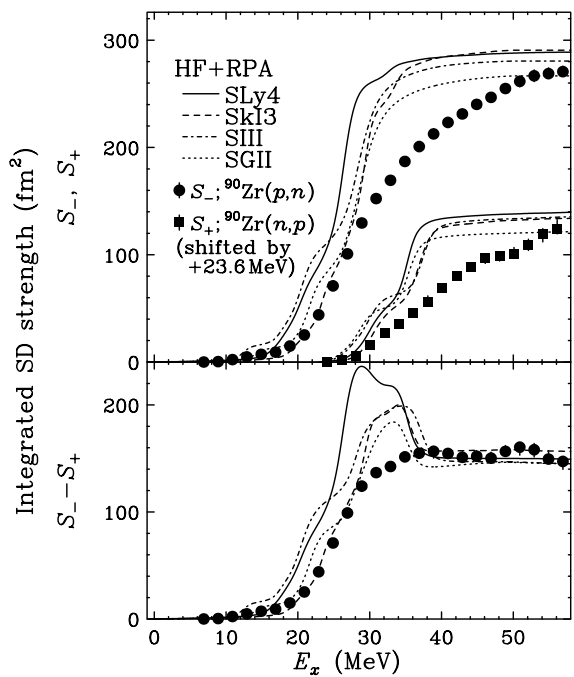}
\caption{\label{fig:zrsum}
Integrated charge exchange SD strength (\ref{eq:sum-ex}) 
excited by  the operators $
\hat{S}_{-} $ and  $
\hat{S}_{+} $ in Eq. (\ref{eq:eq1}) on $^{90}$Zr.  The
calculated results are obtained by the HF+RPA model using 
the Skyrme interactions  SIII,  SGII, SLy4 and  SkI3 \cite{Sagawa07}.
The upper panel shows the $S_-$ and $S_+$ strength, while 
the lower panel shows the  $S_--S_+$ strength.
 All strengths for the 
three multipoles $\lambda^{\pi}$=0$^{-}$, 1$^{-}$ and 2$^{-}$ are
summed up in the results.
The experimental data are taken from Ref. \cite{Yako06}. 
 No quenching 
factor is introduced in the calculation of the integrated strength.
}
 \end{figure}

\begin{table*}[htp]
\caption{\label{tab:sum}
  Sum rule values of charge exchange SD excitations in A=90 nuclei
obtained by the  HF+RPA calculations  \cite{Sagawa07} (S$_-$ for  $^{90}$Nb and
 S$_+$ for  $^{90}$Y).  The SD strength is integrated up to $E_x$ = 50 MeV 
for  S$_-$  and $E_x$ = 26 MeV for S$_+$, respectively. 
The experimental data are taken from Ref. \cite{Yako06}.
The SD sum rules are given in units of fm$^2$.   
See the text for details.}
\footnotesize
\begin{tabular}{l||c|c|c||c|c|c||c|c|c||c|c|c} 
\hline
  &\multicolumn{3}{c||}{SIII} &\multicolumn{3}{c||}{SGII}
  &\multicolumn{3}{c||}{SkI3}&\multicolumn{3}{c}{SLy4} \\ \hline
$\lambda^{\pi} $ & $S_-$ & $S_+$ & $\Delta S$ & $S_-$ & $S_+$ & $\Delta S$
& $S_-$ & $S_+$ & $\Delta S$ & $S_-$ & $S_+$ & $\Delta S$ \\ \hline
 0$^-$  & 34.8  &18.5 &16.4 & 33.2 & 17.4 & 15.8  & 36.6 & 19.1& 17.5&37.8 &21.4 & 16.4\\\hline
 1$^-$  & 120.8 & 71.7 & 49.1 & 122.0 & 74.3 & 47.7 & 120.8  & 68.2 & 52.7 &115.8 & 66.4& 49.4\\\hline
 2$^-$  & 130.1 & 48.5 & 81.6 & 125.5 & 45.9  & 79.5 & 139.0 & 51.1 & 87.9&138.7 & 56.4& 82.3\\\hline
sum   & 285.7   & 138.6 &147.1  & 280.7  &137.6  &143.1   & 296.3 &  138.3   & 158.0 & 292.3 & 144.2 & 148.1\\ \hline \hline 
exp &  \multicolumn{3}{c||}{ $S_-=271\pm14$} &\multicolumn{3}{c||}{
$S_+=124\pm11$ }
  &\multicolumn{3}{c||}{$\Delta S=147\pm13$}&\multicolumn{3}{c}{} \\ \hline
\end{tabular}
\end{table*}

Let us now discuss the integrated SD strength.
The integrated SD strength 
\begin{equation}
 m_0(E_x)=\sum_{\lambda^{\pi}\  =\  0^-,1^-,2^-}\int_0^{E_x}\frac{dB(\lambda^\pi)}{dE'}
dE'
\label{eq:sum-ex}
\end{equation}
 is plotted as a function of 
the excitation energy $E_x$ in Fig. \ref{fig:zrsum} 
for the operators $\hat{S}^{\lambda}_{- } $
and 
 $\hat{S}^{\lambda}_{+} $  in Eq. (\ref{eq:eq1}).
The value $S_-$ is obtained by integrating up to $E_x$ = 50 MeV from the 
ground state of the daughter nucleus $^{90}$Nb ($E_x$ = 57 MeV from 
the ground state of the parent nucleus $^{90}$Zr),
while the corresponding value $S_+$ is evaluated up to $E_x$ = 26 MeV 
from the ground state of $^{90}$Y ($E_x$ = 27.5 MeV from 
the ground state of $^{90}$Zr). 
 This difference between the two maximum energies of the integrals 
 stems from the 
isospin difference 
  between the ground states of the daughter nuclei, i.e.,
 T=4 in $^{90}$Nb and T=6 in $^{90}$Y.
That is, the 23.6 MeV difference originates from the difference in 
excitation energy 
between the T=6 
  Gamow-Teller 
 states in the (p,n) and (n,p)
 channels \cite{Yako06}.
For both the $S_-$ and $S_+$ strength,
the calculated results overshoot the experimental data 
in the energy range $E_x$ = 20-40 MeV.  
  These results suggest a quenching of 
30-40\% of the calculated strength around the peak region.
However, the integrated cross sections up to $E_x$ = 56 MeV in Fig.
 \ref{fig:zrsum}  
approach the calculated values for both the 
$t_{-}$ and $t_{+}$ channels.

The $\Delta S=S_--S_+$ value is shown as a function of $E_x$ 
 in the lower panel of Fig. 
 \ref{fig:zrsum}.  We note that the 
 $\Delta S$ value saturates both in the calculated and the experimental 
 values above $E_x=$ 40 MeV, while the empirical values $S_-$ and $S_+$
 themselves increase gradually above $E_x=$ 40 MeV.  
This is the crucial feature for extracting the model-independent sum rule 
$\Delta S=S_--S_+$ from the experimental data.
The empirical values $S_-$,  $S_+$ and  $\Delta S$ obtained from these 
analyses are shown in Table \ref{tab:sum}. 
   The indicated uncertainties of $S_-$, $S_+$ and $\Delta S$ contain not only
   the statistical error of the data, but also errors due to the various
   input of the DWIA calculations used in the MDA, such as 
   the optical model parameters and the single-particle potentials
   \cite{Yako}. There is an additional uncertainty in the 
   estimation of the SD unit cross section, namely, the overall 
   normalization factor  \cite{Yako06}, which should be studied further 
   experimentally.

\begin{table}[htp]
\caption{\label{tab:hf-zr}
 Proton, neutron and charge radii of $^{90}$Zr.
 The charge radius is obtained by folding the proton density with the proton finite size.
 The sum rule values $\Delta S=S_--S_+$ of spin-dipole excitations are 
  calculated 
 by Eq. (\ref{eq:sum_sd}) with the HF neutron and proton mean square radii. The 
experimental value of the charge radius is taken from Ref. \cite{Vries}, while 
the experimental data for $r_n-r_p$ are taken from \cite{pscatt1,Yako06}. 
 The radii are given in 
units of fm, while the SD sum rules are given in units of fm$^2$.}

\begin{tabular}{l|c|c|c|c|l}   
 \hline
 & SIII & SGII & SkI3 &SLy4 &exp \\ \hline
$r_p$ &  4.257 & 4.198 & 4.174 & 4.225  & 4.19 (from $r_c$)\\
$r_c$  & 4.321 & 4.263  & 4.240 & 4.290 &4.258$\pm$0.008\\
$r_n$ &  4.312  & 4.253 & 4.280 & 4.287& 4.26$\pm$0.04 \cite{Yako06} \\ \hline
$r_n-r_p$ & 0.055 & 0.055 & 0.106 & 0.064 &0.09$\pm$0.07 \cite{pscatt1}  \\
     & & & & & 0.07$\pm$0.04 \cite{Yako06}   \\ \hline
$\Delta S$ & 147.1 & 143.1 & 158.0 & 148.1 &  \\
\hline
\end{tabular}
\end{table}

 From  $\Delta S$, the neutron radius of
$^{90}$Zr is extracted to be $\sqrt{<r^2>_n}$ =  4.26$\pm$0.04 fm 
from the model-independent SD sum rule (\ref{eq:sum_sd}), where the empirical
proton radius \\
$\sqrt{<r^2>_p}$ = 4.19 fm is used. The proton radius 
is obtained from the charge radius in Table \ref{tab:hf-zr} by
 subtracting the
proton finite size correction. 
The experimental uncertainty in the neutron skin thickness 
obtained by proton scattering is rather large: 
   $\delta_{np}=r_n-r_p= 0.09\pm0.07$ fm.  This is 
 mainly due to the difficulty to extract the neutron radius from the  analysis 
of the proton scattering
 ~ \cite{pscatt1}.
 The sum rule analysis of the SD strength 
determines the neutron radius with 1\% accuracy, which is 
almost the same as  that expected for  the parity violation electron 
 scattering experiment.
 The obtained value  $r_n-r_p = 0.07\pm0.04$ fm
  can  be used to disentangle the neutron matter 
 EOS by using the strong linear correlation between the two quantities 
  \cite{Brown:2000,Furnstahl:2002,Yoshi04}.
  
  Very recently, Wakasa performed MDA of $(p,n)$ reaction cross sections on $^{208}$Pb
observed at RCNP, Osaka University and extracted spin-dipole strength in $^{208}$Bi  \cite{Wakasa2013}.  He found 100 \% of the calculated sum rule strength for 1$^-$ states, and about 70 \% 
of the predicted strength for 0$^-$ and 2$^-$ states  \cite{Sagawa07}.   It would be of paramount importance 
to perform the counter experiment $^{208}$Pb$(n,p)$ and extract empirically the model independent sum rule $S_- -S_+$ from the two charge-exchange experiments, in
order to obtain the neutron skin 
value in $^{208}$Pb.

\section{Summary}\label{summary}

In this paper, we have focused on the main constraints on the symmetry energy that are provided by the
experimental and theoretical studies of nuclear collective vibrations. We have not been fully exhaustive on this
subject, in keeping with the fact that other contributions in the present volume deal with the issues we have not
discussed. Thus, our discussion has concerned the isovector dipole and isovector quadupole states, as well
as the isotopic dependence of the isoscalar monpole energies.

It is quite natural to think of the residual proton-neutron force sustaining the isovector collective motion as
being related with the symmetry energy. However, more effort is needed to make this statement more quantitative.
As for the standard GDR, it has been suggested that its energy is correlated with the value of the symmetry
energy at some sub-saturation density around 0.1 fm$^{-3}$, $S(0.1)$ - if medium-heavy nuclei are considered. It is
remarkable that the isovector GQR can be shown to lead to a consistent value of $S(0.1)$. Also, the combination
of $J$ and $L$ that can be deduced is nicely consistent with other kinds of (completely independent)
analysis that are presented in this volume: Eqs.  (\ref{finalcon}), (\ref{l-gqr}) and (\ref{dr-gqr}) substantiate these
statements.

We have also discussed the role played by the pygmy states, or resonances. Empirically, a correlation of their
fraction of EWSR with the slope parameter $L$ has been found, and reasonable values of $L$ (\ref{l-pdr}) and of
the neutron skin (\ref{dr-pdr}) have been extracted. It is puzzling, though, that the PDR does not display in
all the considered models a clear character related to the pure skin mode. This is one of the issues deserving
further investigation.

All these observables do not seem capable of constraining the parameter $K_\tau$, associated with the
second derivative of the symmetry energy. However, a completely different observable namely the dependence
of the isoscalar monopole energy along an isotopic chain, can provide such a constrain. We have discussed 
the theoretical arguments behind that, and the measurements in the Sn and Cd isotopic chains that led
to $K_\tau$ around $-$550 MeV (with a significant error bar still). We have also illustrated the correlations
emerging from our theoretical study between $K_\tau$ and the other parameters associated with the
symmetry energy or, more generally, with the equation of state.

A further source of information on the symmetry energy is the charge-exchange spin-dipole resonance.
In fact, the sum rule obtained from the difference between the total strength in the $t_-$ channel and
the total strength in the $t_+$ channel is proportional to the difference $N\langle r_n^2\rangle -
Z\langle r_p^2\rangle$. Experiments aimed at extracting the neutron skin have been first performed
in the Sn isotopes. More recently, in $^{90}$Zr, it has been possible to extract quantitatively the values
of the total strengths and of the skin. It would be highly desirable to consider the case of $^{208}$Pb as
well, in keeping with the fact that much effort is devoted to the study of this nucleus by using also parity-violating
asymmetry measurements.

In conclusion, the study of giant resonances has been shown to provide some robust conclusions about
the symmetry energy and its density dependence around nuclear matter saturation density. It is not
completely evident how to improve on these first conclusions. Exploring nuclei with larger proton-neutron
asymmetry (unstable nuclei) is of paramount importance as the results may either confirm the present
findings or lead to some surprise. At the same time, further theoretical work is probably needed in order to assess
which correlations with the EoS parameters are genuine, and which are somehow an artefact of a specific
ansatz built in the energy functional.

\section*{Acknowledgements}
This work has been supported in part by the U.S. National Science Foundation (Grant Nos. PHY07-58100 and PHY-1068192).

\end{document}